\documentclass{elsart}

\usepackage{graphicx, amssymb}

\begin{document}

\begin{frontmatter}

\title{Vortex Lattice in Bi$_{2}$Sr$_{2}$CaCu$_{2}$O$_{8+\delta }$ Well Above the
First-Order Phase-Transition Boundary}
\author{J. H. S. Torres\corauthref{cor}},
\ead{henrique@ifi.unicamp.br}
\author{R. Ricardo da Silva\thanksref{lorena}},
\author{S. Moehlecke},
\author{Y. Kopelevich}

\address{Instituto de F\'{\i}sica ``Gleb Wataghin'', Universidade Estadual de\\
Campinas, Unicamp 13083-970, Campinas, S\~{a}o Paulo, Brasil}

\corauth[cor]{Corresponding author.} 
\thanks[lorena]{Also at FAENQUIL, 12600-000, Lorena, SP, Brasil}

\begin{abstract}
Measurements of non-local in-plane resistance originating from transverse
vortex-vortex correlations have been performed on a Bi$_{2}$Sr$_{2}$CaCu$_{2}$O%
$_{8+\delta }$ high-T$_{c}$ superconductor in a magnetic field up to 9 T
applied along the crystal c-axis. Our results demonstrate that a rigid
vortex lattice does exist over a broad portion of the magnetic field --
temperature (H-T) phase diagram, well above the first-order transition
boundary H$_{FOT}$(T). The results also provide evidence for the vortex
lattice melting and vortex liquid decoupling phase transitions, occurring
above the H$_{FOT}$(T).

\end{abstract}


\end{frontmatter}

The knowledge of the magnetic field -- temperature (H-T) phase diagram is a
cornerstone of the phenomenological description of superconductors. The H-T
phase diagram of conventional type-II superconductors is well known. In the
Meissner-Ochsenfeld state, the surface currents screen the applied magnetic
field. Above the lower critical field H$_{c1}$(T) the applied field
penetrates the superconductor in the form of an Abrikosov vortex lattice which
persists up to the upper critical field H$_{c2}$(T), where the
superconductivity vanishes in the bulk of the sample. On the other hand, in
high-temperature superconductors (HTS), due to strong thermal fluctuations,
the occurrence of the vortex lattice melting phase transition at H$_{m}$(T) $%
\ll $ H$_{c2}$(T) has been predicted (for review articles see Ref. \cite{Blatter} and
Ref. \cite{Brandt} and references therein). Since then, considerable efforts have been
dedicated to identifying the melting transition in experiments. At present, it
is widely believed that the vortex lattice melting is related to a first-order 
transition (FOT) occurring, e. g. in Bi$_{2}$Sr$_{2}$CaCu$_{2}$O$_{8}$
(Bi2212) high-Tc superconductor, at a very low field H$_{FOT}$(T) \cite{Zeldov}. Thus,
at T $\thicksim $ T$_{c}$/2 the H$_{FOT}$ $\thicksim $ 500 Oe, which is
several orders of magnitude smaller than H$_{c2}$ $\thicksim $ 100 T. At the
FOT, the equilibrium magnetization jump $\Delta $M$_{eq}$(H,T) takes place
\cite{Zeldov}, which, together with the Clausius-Clapeyron equation, would imply the
occurrence of the entropy jump $\Delta $s(H,T) associated with the
transition. However, an alternative explanation for the magnetization
(entropy) jump due to first-order depinning transition is also possible \cite{Kopel1}.
At this floating-type transition, the vortex lattice does not melt but does the
opposite, it decouples from the atomic lattice becoming more ordered. The
entropy jump associated with the vortex lattice floating transition has
recently been obtained in Monte Carlo simulations by Gotcheva and Teitel
\cite{Gotcheva}. The occurrence of the floating vortex solid phase situated between a
pinned vortex solid and a vortex liquid state has been proposed two decades
ago by Nelson and Halperin \cite{Nelson1}. The FOT in HTS is accompanied by a sudden
increase in the electrical resistivity \cite{Fuchs} which demonstrates the sharpness
of the vortex depinning onset. A sharp resistive magnetic-field-induced
depinning transition separating disordered (low-field) from ordered
(high-field) vortex states has recently been reported for NbSe$_{2}$,
suggesting the first-order nature of the depinning transition \cite{Paltiel}. In Bi2212,
the in-plane non-local resistance, the indication of a finite shear stiffness
of the vortex matter, has been measured above the FOT by Eltsev et al.\cite{Eltsev}.
It has been concluded, however, that the strong transverse vortex-vortex
correlations take place in the vortex liquid phase \cite{Eltsev}.

In the present work, we report the observation of in-plane non-local
resistance in Bi2212 in an applied magnetic field up to H = 9 T. The obtained
results demonstrate that the vortex lattice does exist over a broad portion
of the H-T phase diagram above the FOT boundary.

The resistance measurements were performed on a $l \times w \times t = 1.94 \times%
0.28 \times 0.03 mm^{3}$ size Bi2212 single crystal grown using the self-flux method. The
crystal characterization details as well as dc magnetization measurements
have been presented elsewhere \cite{Kopel2}. The crystal zero-field superconducting
transition temperature T$_{c}^{0}$ = 87.7 K has been determined from the
maximum of the temperature derivative dR/dT. The resistance measurements
were made using PPMS (9T magnet) Quantum Design commercial equipment in
magnetic fields applied along the crystal c-axis. The FOT boundary has
been obtained by means of dc magnetization M(H,T) measurements (H $\parallel 
$ c - axis) with the SQUID magnetometer MPMS5 (Quantum Design).

\begin{figure}
\begin{center}
\includegraphics[scale=.45]{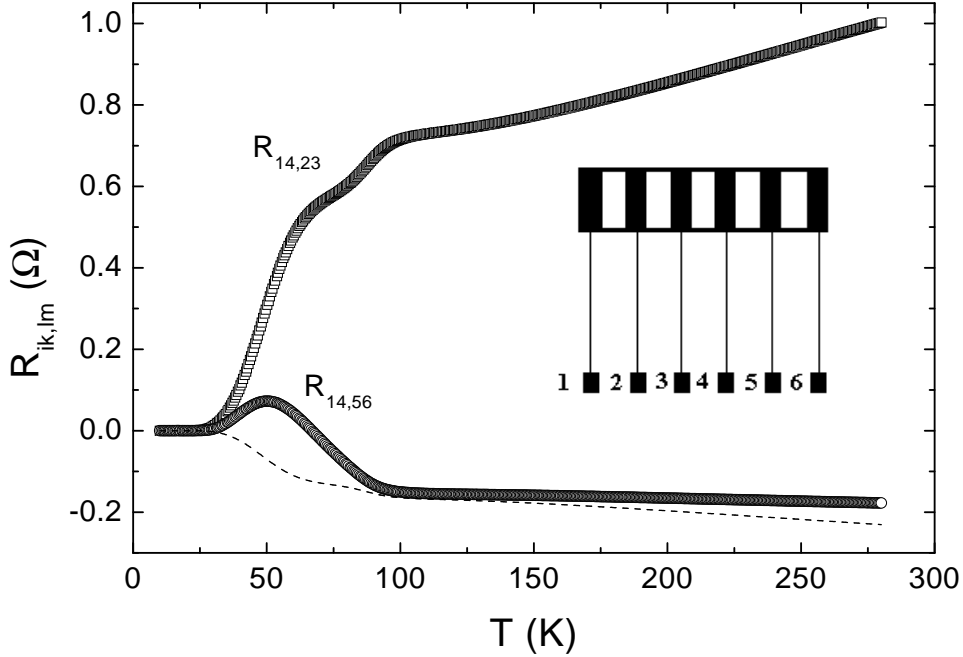}
\end{center}
\caption{Temperature dependences of ``primary'' R$_{14,23}$ = V$_{14,23}$/I$%
_{14}$ and ``secondary'' R$_{14,56}$ = V$_{14,56}$/I$_{14}$ resistances
measured in applied magnetic field H = 9 T and I$_{14}$ = 1 mA; dotted line
is the ``parasitic'' contribution to R$_{14,56}$, resulting from the current
distribution effect as estimated from Eq. (\ref{eqn1}). Inset shows the geometry
of the experiment.}
\label{fig1}
\end{figure}

We used the line-electrode geometry \cite{Mamin,Wortis,Kopel3} to study the non-local in-plane
resistance. Six silver epoxy electrodes with contact resistance $\thicksim $
1$\Omega $ were patterned on one of the main surfaces of the crystal, as
shown in the inset of Fig.\ref{fig1}, with a separation distance s $\thicksim $ 180 $%
\mu $m. In the experiments, the dc current I$_{14}$ was applied between the
current leads 1 and 4, and the voltage was measured simultaneously in both
the applied current part of the crystal V$_{23}$ and outside this region V$_{56}$.

Figure \ref{fig1} shows temperature dependences of both ``primary'' R$_{14,23}$ = V$%
_{14,23}$/I$_{14}$ and ``secondary'' R$_{14,56}$ = V$_{14,56}$/I$_{14}$
resistances obtained in applied field H = 9 T and for I$_{14}$ = 1 mA. As can
be seen from Fig. 1, R$_{14,56}$ is negative in the normal state, and shows
a crossover to positive values below a certain temperature within the
superconducting state.

\begin{figure}
\begin{center}
\includegraphics[scale=.45]{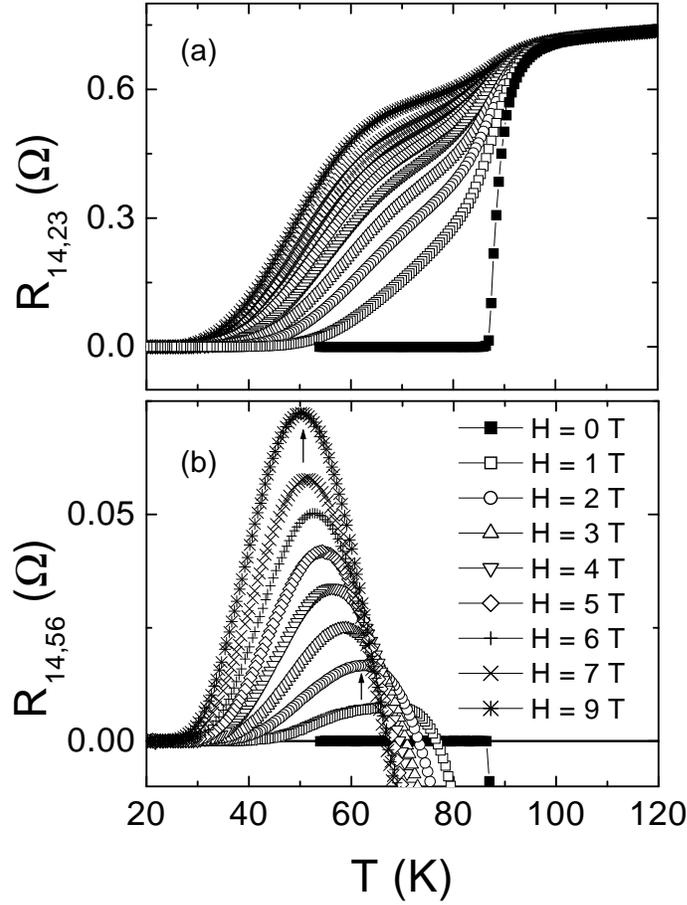}
\end{center}
\caption{``Primary'' resistance R$_{14,23}$(T) (a) and ``secondary''
resistance R$_{14,56}$(T) (b) measured at various applied magnetic fields
and I$_{14}$ = 1 mA. Arrows (b) indicate the field-dependent temperature T$%
_{max}$(H), where a single peak in R$_{14,56}$(T,H) takes place. Symbols in
(a) and (b) correspond to the same fields (b).}
\label{fig2}
\end{figure}

The negative R$_{14,56}$, which develops with the increase of temperature is
related to the current distribution through the crystal thickness, i. e. has
a local origin. Using the equation derived by van der Pauw \cite{Pauw}

\begin{equation}
R_{14,56}=-(wR_{14,23}/\pi s)\ln [(a+b)(b+c)/b(a+b+c)],
\label{eqn1}
\end{equation}
where a, b, and c are distances between electrodes 1 and 4, 4 and 5, 5 and
6, respectively, we obtain the R$_{14,56}$(T), depicted in Fig. \ref{fig1} by a
dotted line. The agreement between calculated and measured R$_{14,56}$(T) is
rather good, taking into consideration the strong crystallographic anisotropy
of Bi2212, the finite width of the electrodes, and a probable distortion of
the electrical potential along the line contacts. Shown in Fig. \ref{fig2}(a) and
Fig. \ref{fig2}(b) are R$_{14,23}$(T,H) and R$_{14,56}$(T,H), respectively measured
at various applied magnetic fields. As can be seen from Fig. \ref{fig2}(b), the
positive ``secondary'' resistance R$_{14,56}$(T,H) emerges and increases
with field. For H $>$ 2 T, the R$_{14,56}$(T,H) shows a well defined peak at
the temperature T$_{max}$(H), which decreases with the field increase.

\begin{figure}
\begin{center}
\includegraphics[scale=.45]{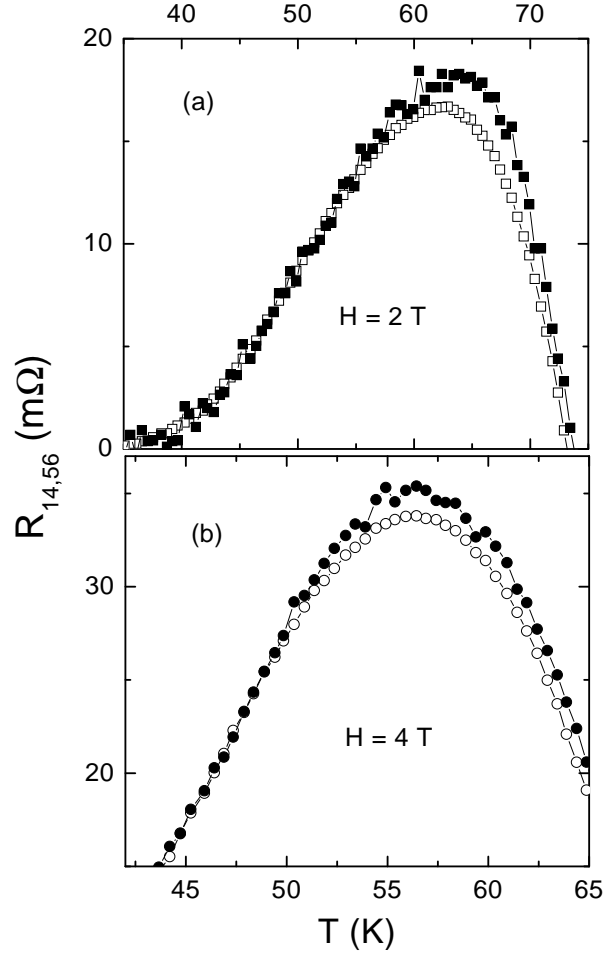}
\end{center}
\caption{``Secondary'' resistance R$_{14,56}$(T) measured with I$_{14}$ = 100
$\mu $A (solid symbols) and I$_{14}$ = 1 mA (open symbols) for H = 2 T (a)
and H = 4 T (b).}
\label{fig3}
\end{figure}

\begin{figure}
\begin{center}
\includegraphics[scale=.45]{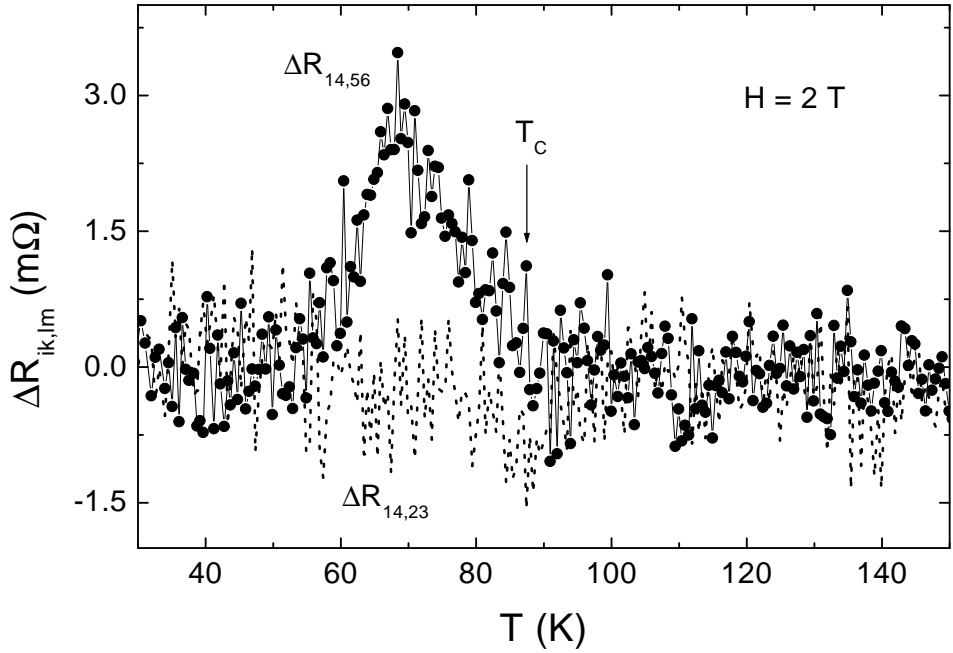}
\end{center}
\caption{The difference $\Delta $R$_{14,56}$ = R$_{14,56}$(I$_{14}$ = 100 $%
\mu $A) -- R$_{14,56}$(I$_{14}$ = 1 mA) (solid symbols) and $\Delta $R$%
_{14,23}$ = R$_{14,23}$(I$_{14}$ = 100 $\mu $A) -- R$_{14,23}$(I$_{14}$ = 1
mA) (dotted line) versus temperature measured for H = 2 T.}
\label{fig4}
\end{figure}

\begin{figure}
\begin{center}
\includegraphics[scale=.45]{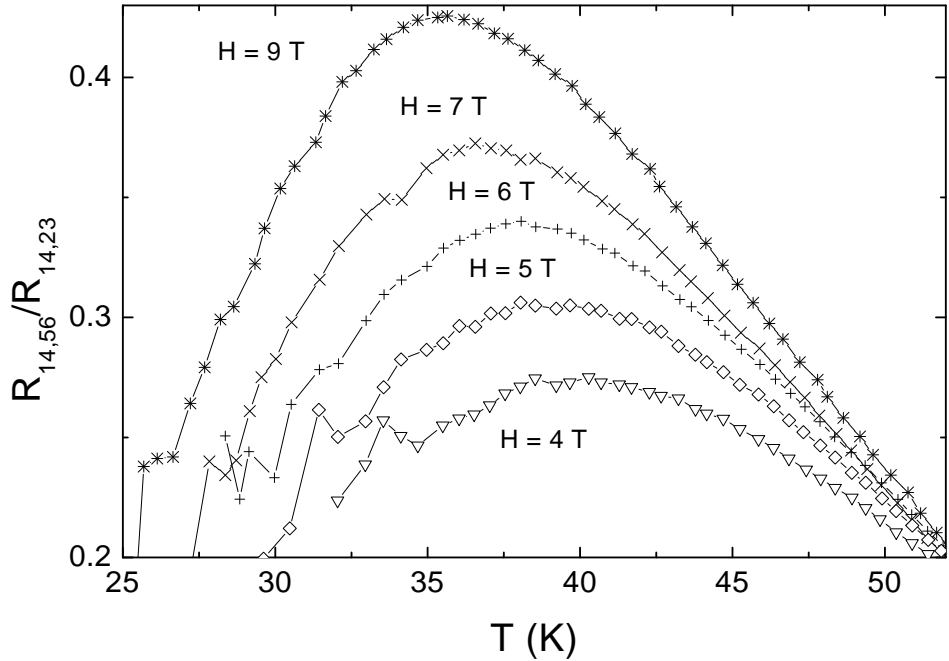}
\end{center}
\caption{The ``secondary'' to ``primary'' resistance ratio R$_{14,56}$/R$%
_{14,23}$ vs. temperature obtained with I$_{14}$ = 1 mA  illustrated for several 
measuring fields.}
\label{fig5}
\end{figure}

The results presented in Figs. \ref{fig3}--\ref{fig5} unambiguously demonstrate that the
positive contribution to R$_{14,56}$(T,H) originates from the non-local
resistance, i. e. is not related to the current distribution effects.

Figure \ref{fig3} (a, b) shows R$_{14,56}$(T) measured with I$_{14}$ = 100 $\mu $A
and I$_{14}$ = 1 mA for H = 2 T [Fig. \ref{fig3}(a)] and H = 4 T [Fig. \ref{fig3}(b)]. In
Fig. \ref{fig4} we plotted $\Delta $R$_{14,56}$ = R$_{14,56}$(I$_{14}$ = 100 $\mu $A)
-- R$_{14,56}$(I$_{14}$ = 1 mA) and $\Delta $R$_{14,23}$ = R$_{14,23}$(I$%
_{14}$ = 100 $\mu $A) -- R$_{14,23}$(I$_{14}$ = 1 mA) versus temperature.
From Fig. \ref{fig3} and Fig. \ref{fig4} we note that (1) as the temperature approaches T$_{max}$%
(H) from below, R$_{14,56}$ becomes larger for smaller applied current,
and the R$_{14,56}$(T,H) peaks at T $\thickapprox $ T$_{max}$ (note that $%
\Delta $R$_{14,23}$ $\thickapprox $\ 0, i. e. R$_{14,23}$ is current-independent 
at the studied temperatures), (2) the current dependence of R$%
_{14,56}$ persists up to $\sim $ T$_{c}$, i. e. is essentially related
to the superconducting state. Note also that the current effect vanishes
with field, so that it is negligible for H $\geq $ 6 T. At T $<$ T$_{max}$%
(H) the ratio R$_{14,56}$/R$_{14,23}$ increases with temperature, as shown in
Fig. \ref{fig5}, for several studied fields. All these experimental facts can hardly
be understood within a local approach, indeed.

Certainly, the large enhancement of the non-local resistance with temperature
over a broad temperature interval cannot be accounted for by vortex-vortex
correlations in the vortex liquid. On the other hand, the long-range
positive non-local resistance can arise from a correlated transverse motion
of the vortex lattice \cite{Wortis}. The in-plane vortex-vortex correlations
occurring on a millimeter scale have also been detected in Corbino-disk
experiments \cite{Lopez,Eltsev}.

The ratio R$_{14,56}$(T)/R$_{14,23}$(T) $<$ 1, see Fig. \ref{fig5}, can result from
the vortex pinning effect which destroys the long-range positional order in the
vortex lattice and, therefore, leads to a depression of the non-local
resistance at large distances. As the temperature increases, both R$_{14,56}$(T)
and R$_{14,56}$(T)/R$_{14,23}$(T) increase due to a vortex pinning
efficiency decrease. Approaching T$_{max}$(H), the ratio R$_{14,56}$(T)/R$%
_{14,23}$(T) starts to decrease.

\begin{figure}
\begin{center}
\includegraphics[scale=.45]{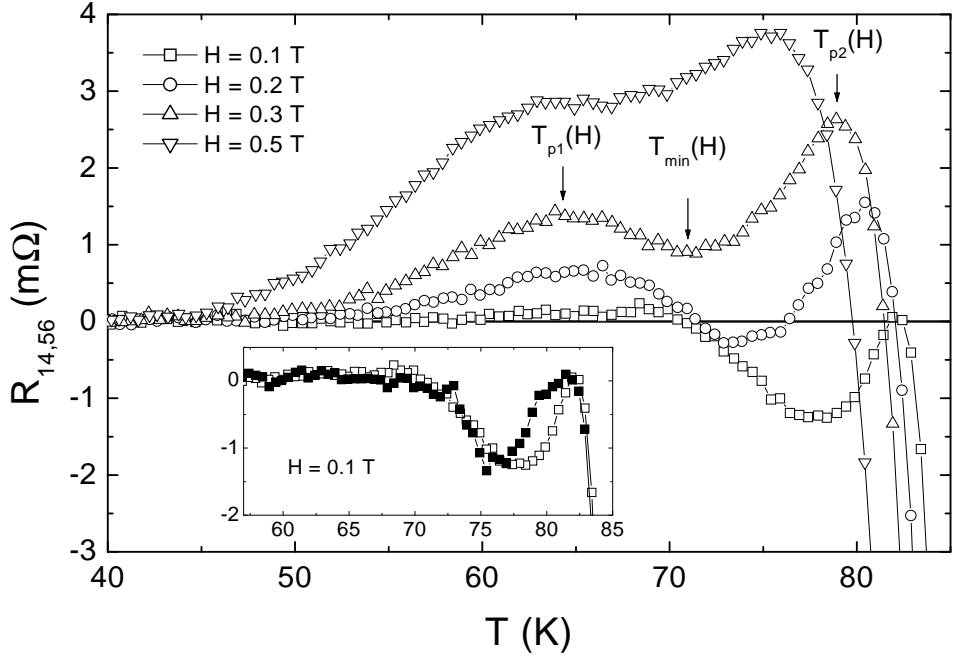}
\end{center}
\caption{The ``secondary'' resistance R$_{14,56}$(T) measured for several low
(see text) fields with I$_{14}$ = 1 mA. The inset shows R$_{14,56}$(T)
measured at H = 0.1 T with I$_{14}$ = 100 $\mu $A (solid symbols) and I$_{14}
$ = 1 mA (open symbols).}
\label{fig6}
\end{figure}

\begin{figure}
\begin{center}
\includegraphics[scale=.45]{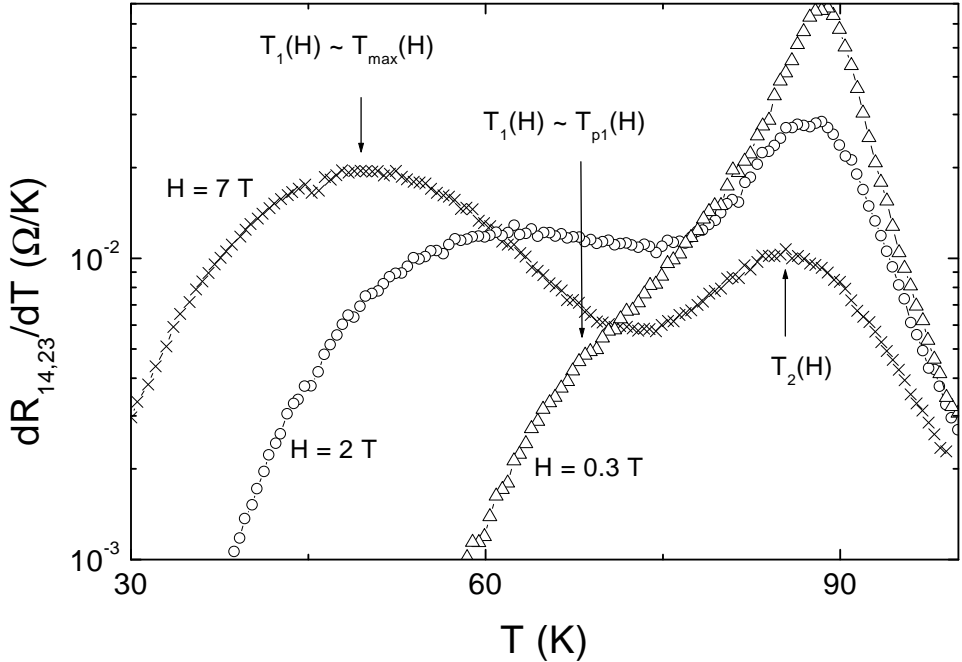}
\end{center}
\caption{The derivative dR$_{14,23}$/dT vs. T demonstrating the slowing-down
of the ``primary'' resistance increase at T $\geq $ T$_{1}$(H) which is
situated just below the T$_{max}$(H) at high fields (H $\geq $ 2 T) and
coincides with the T$_{p2}$(H) at low fields (H $\leq $ 0.5 T). At the
temperature T$_{2}$(H), transition to the normal state takes place.}
\label{fig7}
\end{figure}

On the other hand, the occurrence of the maximum in $\Delta $R$_{14,56}$(T)
at T $\thickapprox $\ T$_{max}$(H), see Fig. \ref{fig4}, rules out any trivial 
(i.e. within a local approach) explanation of the R$_{14,56}$(T) reduction 
above T$_{max}$(H). A non-trivial
origin of the maximum in R$_{14,56}$(T,H) is supported by the observation of
a splitting of this maximum into two peaks which takes place below $\sim $ 1
T, see Fig. \ref{fig6}. The occurrence of two peaks (or hollow) in R$_{14,56}$(T,H)
at low fields can be understood assuming the ``reentrant'' enhancement of
the vortex pinning efficiency in the temperature interval T$_{p1}$(H) $<$ T $%
<$ T$_{p2}$(H), which resembles a phenomenon known as ``peak effect'' (PE)
\cite{Pippard,Larkin}. In agreement with the PE occurrence, the increase of the
``primary'' R$_{14,23}$(T,H) resistance slows down at T $\geq $ T$_{1}$ $%
\sim $ T$_{p1}$, as Fig. \ref{fig7}, where dR$_{14,23}$(T,H)/dT vs. T is plotted, 
illustrates for H = 0.3 T. A similar phenomenon takes place at the
temperature T$_{1}$(H) just below T$_{max}$(H) in the high-field limit, see
Fig. \ref{fig7} (the second peak in the derivative dR$_{14,23}$(T,H)/dT occurring at
temperature T$_{2}$(H) is related to the superconductor-normal metal
transition). In the low-field limit, we have also observed a current-induced
suppression of R$_{14,56}$(T,H) at T $>$ T$_{min}$(H), see inset in Fig. \ref{fig6}.
This implies a similar vortex state occurring above T$_{max}$(H) (high
fields) and above T$_{min}$(H) (low fields).

\begin{figure}
\begin{center}
\includegraphics[scale=.45]{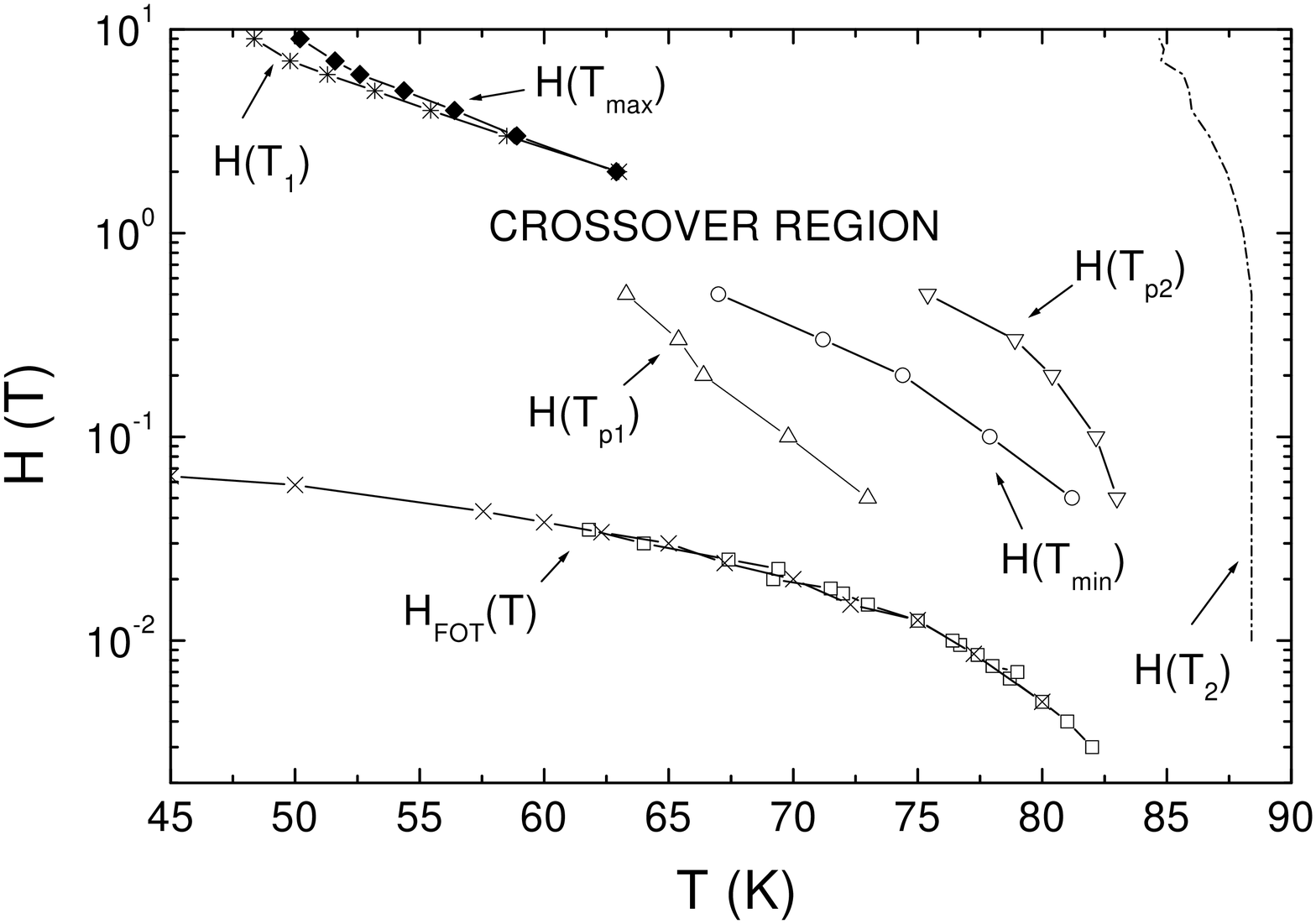}
\end{center}
\caption{Magnetic field -- temperature (H-T) diagram constructed on the base
of experimental results (see text). The first order transition boundary H$%
_{FOT}$(T) measured in both this work ($\Box $) and Ref. [7] (x) is also
shown.}
\label{fig8}
\end{figure}

The above results are summarized in the magnetic field -- temperature (H-T)
diagram (Fig. \ref{fig8}) which we discuss now.

\begin{figure}
\begin{center}
\includegraphics[scale=.45]{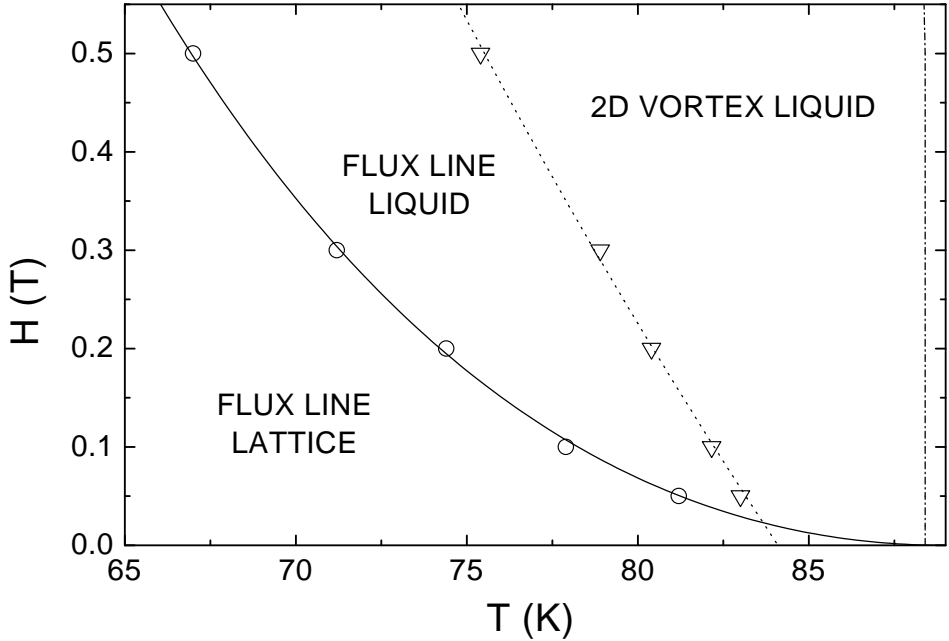}
\end{center}
\caption{Low-field (H $<$ H$_{3D-2D}$) portion of the H-T phase diagram. The
solid line corresponds to the equation $H_{m}(T)=B(1-t^{2})^{2}/t^{2}$, which
describes the flux-line-lattice melting phase transition (Eq. (\ref{eqn3})); B
= 1.5 T, t = T$_{m}$/T$_{c0}$, T$_{c0}$ = 89 K. The dotted line corresponds
to the equation $H_{D}(T)=C(T_{c}-T)/T$ (Eq. (\ref{eqn4})), which describes the
3D-2D ``decoupling'' transition in the vortex liquid state; C = 4.4 T, and T$%
_{c}$ = 84.1 K.}
\label{fig9}
\end{figure}

The FOT boundary obtained by means of dc magnetization measurements (not
presented here) is plotted in Fig. \ref{fig8} together with data from Ref. \cite{Fuchs},
measured for a similar Bi2212 crystal. It is evident from Fig. \ref{fig8} that at T $%
\sim $ T$_{c}$/2, the H$_{FOT}$(T) is about 100 times smaller than the H$%
_{max}$(T), implying that the long-range transverse vortex-vortex
correlations persist well above the FOT boundary. This fact has a natural
explanation, assuming that the floating (depinning) transition is associated with the
FOT \cite{Kopel1,Gotcheva,Nelson1}. At H $<$ H$_{FOT}$(T) $\sim $ $\Phi _{0}$/$\lambda _{ab}^{2}$, the
vortex lattice shear modulus decreases exponentially with field $%
c_{66}\thickapprox (\varepsilon _{0}/\lambda _{ab}^{2})(H\lambda
_{ab}^{2}/\Phi _{0})^{1/4}exp[-(\Phi _{0}/H\lambda _{ab}^{2})^{1/2}]$, whereas at
H $>$ H$_{FOT}$(T), $c_{66}\thickapprox (\varepsilon _{0}/4\Phi _{0})H$, i.
e. c$_{66}$ linearly increases with field \cite{Blatter,Brandt}, where $\varepsilon
_{0}=(\Phi _{0}/4\pi \lambda _{ab})^{2}$. This implies that, at H $<$ H$_{FOT}
$(T), the interaction between vortices and the quenched disorder overwhelms
the vortex-vortex interaction, leading to a stronger vortex pinning in the
low-field regime (note that at H $<$$<$ H$_{FOT}$(T) the pinned vortex
liquid is expected \cite{Nelson2}). With the field increase the c$_{66}$(H,T) and
hence the inter-vortex interaction increase, and the vortex lattice
de-couples from the atomic lattice at H$_{FOT}$(T). Usually, the depinning
transition in HTS is rather sharp \cite{Fuchs}. We stress that besides theoretical
expectations of a sharp depinning transition \cite{Koshelev,Wagner}, an experimental
evidence of a jumpy-like magnetic-field-induced floating transition has
recently been reported \cite{Paltiel}. At H $>$ H$_{FOT}$(T) and low enough
temperatures, c$_{66}$(H,T) is weakly temperature-dependent. In this regime,
the vortex lattice becomes more ordered when the temperature is increased due to
the suppression of the vortex pinning efficiency by thermal fluctuations,
resulting in the increase of the non-local in-plane resistance with
temperature. This observation is in excellent agreement with the
second-order diffraction in small-angle neutron scattering experiments \cite{Forgan}
which revealed the formation of a more ordered vortex lattice with the
temperature increase for intermediate temperatures and magnetic fields.
The c$_{66}$(H,T) rapidly decreases, however, approaching either the upper
critical field H$_{c2}$(T) or the melting phase transition boundary H$_{m}$%
(T) $<$ H$_{c2}$(T). In both cases, the vortex lattice can better adjust the
pinning potential \cite{Pippard,Larkin} leading to the reduction of the non-local signal.
There are two plausible scenarios which allow us to account for the occurrence
of the minimum in R$_{14,56}$(T,H) in the low-field regime, see Fig. \ref{fig6}. The
first possibility is that thermal fluctuations smear out the pinning
potential, improving the vortex lattice which leads to the reentrant increase
of R$_{14,56}$(T,H) with the temperature increase at T $>$ T$_{min}$(H)
\cite{Kopel4}. At T = T$_{p2}$(H), the vortex lattice melts or the superconducting order
parameter diminishes because of strong fluctuations in its amplitude; both
effects will suppress the non-local resistance. On the other hand, the
minimum in R$_{14,56}$(T,H) occurring at the T$_{min}$(H) can coincide with
the melting transition temperature T$_{m}$(H) in the presence of quenched
disorder \cite{Bhattacharya,Tang,Granato}. Then, in a narrow temperature interval above the T$_{m}$%
(H), a shear viscosity due to a finite crossing energy U$_{\times}$(H,T) of the
entangled vortex liquid \cite{Nelson3,Carraro} can lead to the restoration of the
non-local resistance. As temperature increases further, the U$_{\times}$(H,T)
vanishes \cite{Carraro}, and the non-local resistance will be suppressed together with
the entangled vortex state at the ``decoupling'' transition temperature T$%
_{D}$(H) = T$_{p2}$(H) $>$ T$_{m}$(H), above which vortex fluctuations have a
two-dimensional (2D) character. This second scenario agrees with the observed suppression of 
R$_{14,56}$(H,T) at T $>$ T$_{min}$(H) by the applied
current, assuming the occurrence of current-induced vortex cutting \cite{Safar}.
There is also a striking correspondence between the experimental results,
see Fig. \ref{fig9}, and the low-field portion of the H-T phase diagram proposed by
Glazman and Koshelev \cite{Glazman} for layered superconductors. Indeed, the H(T$_{min}
$) can be described perfectly by a theoretical 3D ``melting line'' \cite{Blatter,Brandt,Glazman}

\begin{equation}
H_{m}(T)\cong \Phi _{0}\varepsilon _{0}^{2}c_{L}^{4}/(k_{B}T)^{2}\gamma ^{2},
\label{eqn2}
\end{equation}

where $\gamma $=$\lambda _{c}$/$\lambda _{ab}$ is the anisotropy factor, $%
\lambda _{c}$ is the out-of-plane penetration depth, and c$_{L}$ = 0.1 -- 0.4 is
the Lindemann number. The Eq. (\ref{eqn2}) can be re-written in the form

\begin{equation}
H_{m}(T)=B(1-t^{2})^{2}/t^{2},  \label{eqn3}
\end{equation}

where t $\equiv $\ T$_{m}$/T$_{c0}$, T$_{c0}$ is the mean-field transition
temperature, and $B=\Phi _{0}^{5}c_{L}^{4}/256\pi
^{4}(k_{B}T_{c0})^{2}\gamma ^{2}\lambda _{ab}^{4}(0)$. The fitting gives B =
1.5 T (see Fig. \ref{fig9}). Taking a dimensional crossover field \cite{Glazman} for our
crystal H$_{3D-2D}$ $\cong $\ $\Phi _{0}$/($\gamma $d)$^{2}$ $\sim $ 0.5 T
which separates 3D (H $<$ H$_{3D-2D}$) and quasi-2D (H $>$ H$_{3D-2D}$)
vortex fluctuation regimes, we obtain $\gamma $ $\thickapprox $ 40 (here d = 15 \AA {} 
is the distance between weakly coupled CuO$_{2}$ bi-layers). Then, with $%
\lambda _{ab}$(0) $\sim $ 1000 \AA, one gets a reasonable value for the
Lindemann number c$_{L}$ $=$ 0.23.

On the other hand, H(T$_{p2}$) can be best approximated by the linear
dependence (see Fig. \ref{fig9})

\begin{equation}
H_{D}(T)=C(T_{c}-T)/T,  \label{eqn4}
\end{equation}

which describes the thermally induced 3D-2D vortex liquid decoupling
transition in a vicinity of T$_{c}$ \cite{Glazman}. Here $C=\alpha _{D}\Phi
_{0}^{3}/dk_{B}T_{c}(4\pi \lambda _{c})^{2}$, and $\alpha _{D}$ is some
constant. With the fitting parameter C = 4.4 T, one has $\alpha
_{D}\thicksim 1$. The apparent crossing of H$_{m}$(T) and H$_{D}$(T) lines
seen in Fig. \ref{fig9} originates from the entering into the critical
superconducting fluctuations region (see, e. g., Ref. \cite{Nguyen}).

\begin{figure}
\begin{center}
\includegraphics[scale=.45]{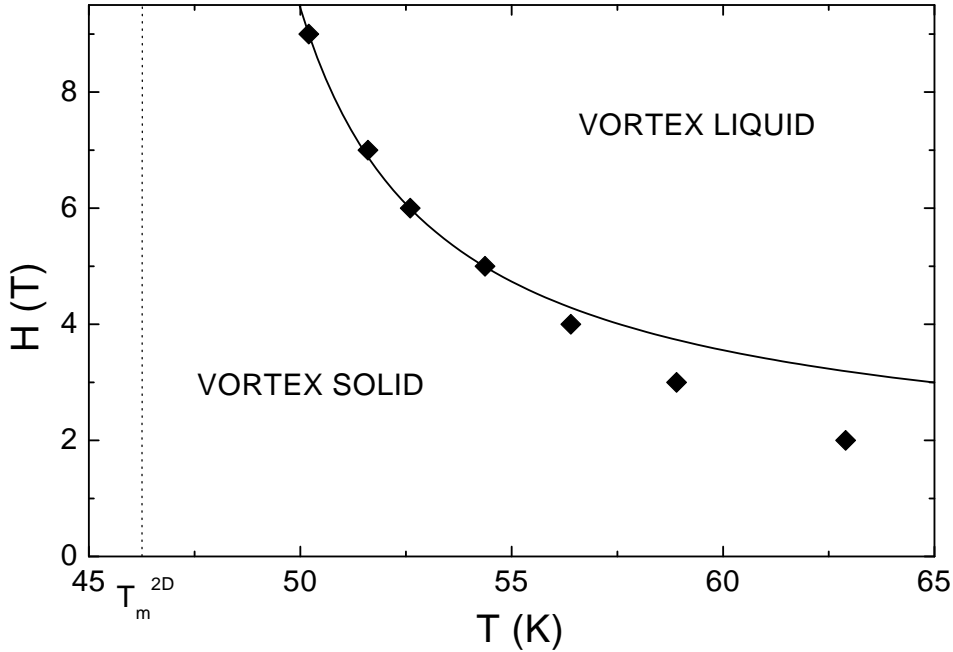}
\end{center}
\caption{The high-field (H $>$ H$_{3D-2D}$) portion of the H-T phase
diagram; ($\blacklozenge $) - H(T$_{max}$), solid line is obtained from Eq. (%
\ref{eqn5}) with H$_{3D-2D}$ = 0.74 T, and T$_{m}^{2D}$ = 46.3 K.}
\label{fig10}
\end{figure}

For H $>$ 0.5 T, H($T_{p1}$) and H(T$_{p2}$) start to merge and, for H $>$ 2
T, a single transition in the vortex matter takes place at T$_{max}$(H). For
H $>$$>$ H$_{3D-2D}$, the theory \cite{Glazman} predicts that H$_{m}$(T) approaches the
melting temperature of an isolated superconducting CuO$_{2}$ bi-layer $%
T_{m}^{2D}\cong (k_{B}8\pi \sqrt{3})^{-1}d\varepsilon _{0}$ according to the
equation:

\begin{equation}
H_{m}(T)\cong H_{3D-2D}exp\{b[T_{m}^{2D}/(T-T_{m}^{2D})]^{\nu }\},
\label{eqn5}
\end{equation}
where b $\sim $ 1, and $\nu $ = 0.37. Figure \ref{fig10} demonstrates a good
agreement between Eq. (\ref{eqn5}) and the experimental H(T$_{max}$) boundary
at H $>$ 4 T. The fitting gives H$_{3D-2D}$ = 0.74 T and T$_{m}^{2D}$ = 46.3
K ($\lambda _{ab}$(0) $\thickapprox $\ 1200 \AA ).

Thus, taking the overall data together, we are led to conclude on a possible
occurrence of vortex lattice melting and ``decoupling'' phase transitions
associated with H(T$_{min}$) and H(T$_{p2}$) low-field boundaries,
respectively, as well as on the melting of a quasi-2D vortex solid which
takes place along the H(T$_{max}$) boundary at $H\gg H_{3D-2D}$. The current
effect on the non-local resistance measured at both T $\geq $\ T$_{min}$(H)
and T $\geq $\ T$_{max}$(H), and its vanishing with the field increase,
suggests the ocurrence of the entangled vortex liquid for low and
intermediate fields. We stress that the results obtained here suggest an
enhancement of the vortex pinning in the vortex liquid state, being in
agreement with Refs. \cite{Bhattacharya,Tang,Granato,Glazman}.

To summarize, results of the present work provide an experimental evidence
for the vortex lattice existence in Bi$_{2}$Sr$_{2}$CaCu$_{2}$O$_{8}$ well
above the first-order transition boundary H$_{FOT}$(T). For the first time,
the H-T phase diagram of the high-T$_{c}$ superconductor is constructed on
the basis of direct probe of transverse vortex-vortex correlations.

We gratefully acknowledge valuable discussions with G. Carneiro, P. Esquinazi, 
E. Granato, V. M. Vinokur, and E. Zeldov.

This work was supported by FAPESP, CNPq and CAPES Brazilian agencies.

\end{document}